\documentclass[12pt,a4paper,dvips]{article}
\usepackage{a4p}
\usepackage{cite,mcite}
\usepackage{graphicx}
\usepackage{physics }
\usepackage{epsfig }
\usepackage{l3_title,ifthen}
\usepackage{Lep}
\Lep{2}
\journalname{Phys. Lett. B}
\date{May 19, 2000}
\preprint{2000-066}
\newlength{\capwidth}
\newcommand{\EE}{\mathrm{e}^+\mathrm{e}^-}
\newcommand{\MM}{\mu^+\mu^-}
\newcommand{\TT}{\tau^+\tau^-}
\newcommand{\LL}{\ell^+\ell^-}
\newcommand{\FF}{\mathrm{f} \, \bar{\mathrm{f}}^\prime}
\newcommand{\QQ}{\mathrm{q} \, \bar{\mathrm{q}}^\prime}
\newcommand{\qq}{\mathrm{q} \, \bar{\mathrm{q}}}
\newcommand{\q} {\mathrm{q}}
\newcommand{\WW}{\mathrm{W}^+\mathrm{W}^-}
\newcommand{\Wen}{\mathrm{e^+} \nu_\mathrm{e} \mathrm{W^-}}
\newcommand{\Wln}{\mathrm{W^-} \to \mathrm{\ell^-} \bar{\nu}_\ell}
\newcommand{\enu}{\mathrm{e^+} \nu_\mathrm{e}}
\newcommand{\ean}{\mathrm{e^-} \bar{\nu}_\mathrm{e}}
\newcommand{\enw}{\mathrm{e} \nu_\mathrm{e} \mathrm{W}}

\begin{document}
\begin{titlepage}

\title{Production of Single W Bosons at
       \bf{\boldmath${\sqrt{s} = 189}$ Ge$\kern -0.1em $V}\\
       and Measurement of \boldmath$\rm W W \gamma$ Gauge Couplings }

\author{The L3 Collaboration}

\begin{abstract}
  
  Single W boson production in electron-positron collisions is
  studied with the L3 detector at LEP. The data sample collected
  at a centre-of-mass energy of $\sqrt{s} = 188.7\GeV{}$ corresponds
  to an integrated luminosity of $176.4\,\pb$. Events with a single
  energetic lepton or two acoplanar hadronic jets are selected.
  Within phase-space cuts, the total cross-section is measured to
  be $0.53 \pm 0.12 \pm 0.03 \; \rm{pb}$, consistent with the
  Standard Model expectation. Including our single W boson results
  obtained at lower $\sqrt{s}$, the WW$\gamma$ gauge couplings
  $\kappa_\gamma$ and $\lambda_\gamma$ are determined to be
  \mbox{$\kappa_\gamma = 0.93 \pm 0.16 \pm 0.09$} and
  \mbox{$\lambda_\gamma = -0.31^{+0.68}_{-0.19} \pm 0.13$}.

\end{abstract}

\submitted

\end{titlepage}


\section {Introduction}

Precise measurements of trilinear gauge boson couplings constitute
a crucial test of the Standard Model of electroweak
interactions~\cite{gsw,veltman}. The studies of single W
production\footnote{The charge conjugate reactions are understood
to be included throughout the paper.}
\begin{equation}
 \EE \to \Wen
 \label{eq:eqn01}
\end{equation}
by the LEP experiments~\cite{l3wen1,l3wen2,alephwen,delphiwen}
demonstrated that this process provides one of the best experimental
grounds for precision measurements of the electromagnetic gauge couplings
of the W boson. The cross-section of process (\ref{eq:eqn01}) depends only
on the $\kappa_\gamma$ and $\lambda_\gamma$ gauge coupling
parameters~\cite{tsukamoto} which are related to the magnetic dipole
moment, $\mu_\mathrm{W} = (e/(2\,\MW))
\left(1 + \kappa_\gamma + \lambda_\gamma \right)$,
and the electric quadrupole moment, 
$\mathrm{Q}_\mathrm{W} = (-e/\MW^2)
\left(\kappa_\gamma - \lambda_\gamma \right)$, of the W boson.
Any deviation from the Standard
Model predictions $\kappa_\gamma = 1$ and $\lambda_\gamma = 0$ would
indicate that the W boson has an internal structure.

A specific feature of this reaction is a final state positron
produced at very low polar angle and therefore not
detected. Thus the signature of this process is a single energetic
lepton, if the $\mathrm{W}^-$ boson decays into lepton and 
anti-neutrino, or two hadronic jets and large transverse
momentum imbalance in case of hadronic $\mathrm{W}^-$ decay.

In this article we report on the measurement of the cross-section of
single W boson production at a centre-of-mass energy of
$\sqrt{s}=188.7\GeV$, denoted hereafter as $\sqrt{s}=189\GeV$.
Combining these results with those on single W boson production
obtained at lower $\sqrt{s}$~\cite{l3wen2}, we derive
significantly more precise values for the gauge couplings
$\kappa_\gamma$ and $\lambda_\gamma$.


\section {Data and Monte Carlo Samples}

The data were collected by the L3 detector~\cite{l3det} at LEP in 1998
with an integrated luminosity of $176.4\,\pb$.

For signal studies samples of $\EE \to \enu \FF$ events are generated
using both the GRC4F~\cite{grc4f} and the EXCALIBUR~\cite{excalibur} 
Monte Carlo generators. For the background studies the following
Monte Carlo programs are used: 
KORALW~\cite{koralw} ($\EE \to \WW \to {\rm f\/f\/f\/f}$),
KORALZ~\cite{koralz} ($\EE \to \MM(\gamma), \; \TT(\gamma)$),
BHAGENE3~\cite{bhagene} and BHWIDE~\cite{bhwide} for large
angle Bhabha scattering ($\EE \to \EE(\gamma)$),
TEEGG~\cite{teegg} for small angle Bhabha scattering 
($\EE \to \EE \gamma$), 
PYTHIA~\cite{pythia} ($\EE \to \qq(\gamma)$), 
DIAG36~\cite{diag36} and PHOJET~\cite{phojet} for leptonic and
hadronic two-photon processes, respectively, 
and EXCALIBUR and GRC4F for other 4-fermion final states.

The Monte Carlo events are simulated in the L3 detector using the
GEANT~3.15 program~\cite{geant}, which takes into account the effects
of energy loss, multiple scattering and showering in the detector. The
GHEISHA program~\cite{gheisha} is used to simulate hadronic
interactions in the detector.


\section {Signal Definition}

The signal definition used here is unchanged with respect to our previous
publications~\cite{l3wen1,l3wen2}. The signal is defined as $\EE \to \enu \,\FF$
events that satisfy the following phase-space cuts:
\parbox{\textwidth}{
\begin{eqnarray}
  | \, \rm{\cos\theta_{e^+}} |        & > & 0.997             \nonumber   \\
  \label{eq:eqn02}
  \mbox{min}(E_\mathrm{f},E_{\bar{\mathrm{f}}^\prime}) & > & 15   \GeV{}  \\ 
  | \, \rm{\cos\theta_{e^-}} |        & < & 0.75  \;\;\;\;\; \rm{for} \;
\enu \ean \; \rm{events \; only,}                             \nonumber
\end{eqnarray}
             }
where $\rm{\theta_{e^+}}$ is the polar angle of the
outgoing positron, and $E_\mathrm{f}$ and
$E_{\bar{\mathrm{f}}^\prime}$ are the fermion energies. The final
states $\rm \EE \to \enu \, \FF$ that do not satisfy these conditions
are considered as background; they consist mostly of the reaction $\EE
\to \WW$. In the case of the $\enu\,\ean$ final state the additional
angular cut reduces contributions from processes where the
$\nu_\mathrm{e}\bar{\nu}_\mathrm{e}$ pair originates from the decay of
a Z boson.

Inside the phase-space region (\ref{eq:eqn02}), single W production 
dominates since it peaks strongly at $\rm | \cos
\theta_{e^+} | \sim 1$. On average it accounts for 90\% of all events
in this region, the remaining 10\% being mostly non-resonant final
states. The purity depends slightly on the flavour of the $\FF$ pair
from $\rm W^-$ decays. For the $\enu \, \ean$ final state, the purity
is 75\%.

For comparison with theory, the cross-sections for this signal definition
are calculated with a statistical precision from 0.2\% to 1.0\%
using the Monte Carlo generators GRC4F and EXCALIBUR. 
The main difference between the two generators is the approximation of
massless fermions used in EXCALIBUR. 
The reduction of theoretical uncertainties on predictions for single W
production is subject to ongoing theoretical efforts~\cite{passarino}. 
With respect to this discussion, we estimate
the theoretical uncertainty on the calculated cross-section
for single W production to be of the order of  7\%. This includes
the effect of using a smaller electromagnetic coupling accounting
for the low momentum transfer of the photon in single W production
and taking into account QED radiative corrections expected for such
a $t$-channel process.
Rescaling with respect to a smaller electromagnetic coupling constant
would lower the cross section by 7\% to 10\%, whereas replacing the
energy scale used in the structure function approach to calculate
QED corrections with the correct physical scale is expected to increase
the cross-section by about 5\%.


\section {Analysis}

All decay modes of the W boson are considered, leading to the following
experimental signatures: events with two hadronic jets and large transverse
momentum imbalance and events with single energetic electrons, muons
and taus. In the following, efficiencies are quoted with respect to
the phase-space cuts (\ref{eq:eqn02}), with errors due to Monte Carlo
statistics.


\subsection {Hadronic Final States}

The selection of candidates for the hadronic decay $\Wm \to \QQ$ of
single W boson production is based on the following requirements: two
acoplanar hadronic jets, no leptons, and large transverse momentum
imbalance.

High multiplicity hadronic events with at least five charged tracks
are selected with an energy deposition in the electromagnetic
calorimeter greater than $10\GeV{}$ and large total visible
energy, $E_\mathrm{vis} > 60\GeV{}$. All calorimetric clusters in 
the event are forced to two hadronic jets using the DURHAM jet finding
algorithm~\cite{durham}. The invariant mass of the jet-jet system must
be in the range $40\GeV{} < M_\mathrm{vis} < 110\GeV{}$. The energy
$E_{\rm FB}$ in the forward-backward luminosity calorimeters, covering
the angular range $0.025\,\mbox{rad} < \theta < 0.151\,\mbox{rad}$,
where $\theta$ is measured with respect to the incoming electron or
positron, is required to be smaller than $65\GeV{}$. These cuts reduce
contributions from the purely leptonic
final states $\EE \to \EE (\gamma), \, \MM (\gamma), \, \TT (\gamma)$
and hadronic two-photon interactions while
keeping a significant fraction of hadronic events from the processes
$\EE \to {\rm q \bar{q}} (\gamma)$, $\EE \to \WW$ and $\EE \to \rm{ZZ}$.

To reject events from the two-fermion production process 
$\EE \to {\rm q \bar{q}} (\gamma)$, the missing transverse momentum must
exceed $15\GeV{}$. The missing momentum vector must be at least
0.30~rad away from the beam axis and the energy in the $\pm 0.22\,\mbox{rad}$
azimuthal sector around its direction must be below $10\GeV{}$. In
addition, the opening angle between the two jets in the plane
perpendicular to the beam axis must be smaller than 3.0~rad.

Events containing identified electrons, muons or photons with energy
greater than $20\GeV{}$ are rejected in order to suppress the
remaining background from $\EE \to \WW$ events where one of the W
bosons decays into leptons. In order to remove part of the remaining
$\tau^+\nu_\tau \, \QQ$ final states with the $\tau$ lepton decaying
hadronically, the event is forced to a three-jet topology
using the DURHAM algorithm. The solid angle, $\Omega$, defined by
the directions of these jets is required to be smaller than 5.5~sr.

At $189\GeV{}$ centre-of-mass energy, the production of two
on-mass-shell Z bosons is a source of an additional background. A
decay of one Z into neutrinos accompanied by a hadronic decay
of the other Z leads to an event signature close to that of the signal
process (\ref{eq:eqn01}). To suppress this background, we make use of
the fact that the Z bosons produced in pairs are close to the
kinematic threshold and therefore have low momenta. Thus we ask the
velocity of the detected hadronic system, $\beta $, calculated as the
ratio of the missing momentum to the visible energy to be greater
than 0.35.

A total of 216 events are observed in the data, with 35.9 events expected
from the signal according to the EXCALIBUR Monte Carlo prediction and 179
from the background sources distributed as follows: $\WW$ production
and non resonant $\enu \QQ$ final states (166.2 events), ZZ 
production (7 events), two-fermion final state processes (4.2 events) and
two-photon interactions (1.5 events). The contributions to the
background from other processes are found to be negligible. The
number of observed events is in good agreement with the expectation.
The signal selection efficiency is determined to be $(50.5 \pm 0.7)$\%
using the EXCALIBUR Monte Carlo and $(51.8 \pm 1.3)$\% using GRC4F Monte
Carlo.

\begin{figure} [ht]
  \begin{center}
    \mbox{\epsfig{file=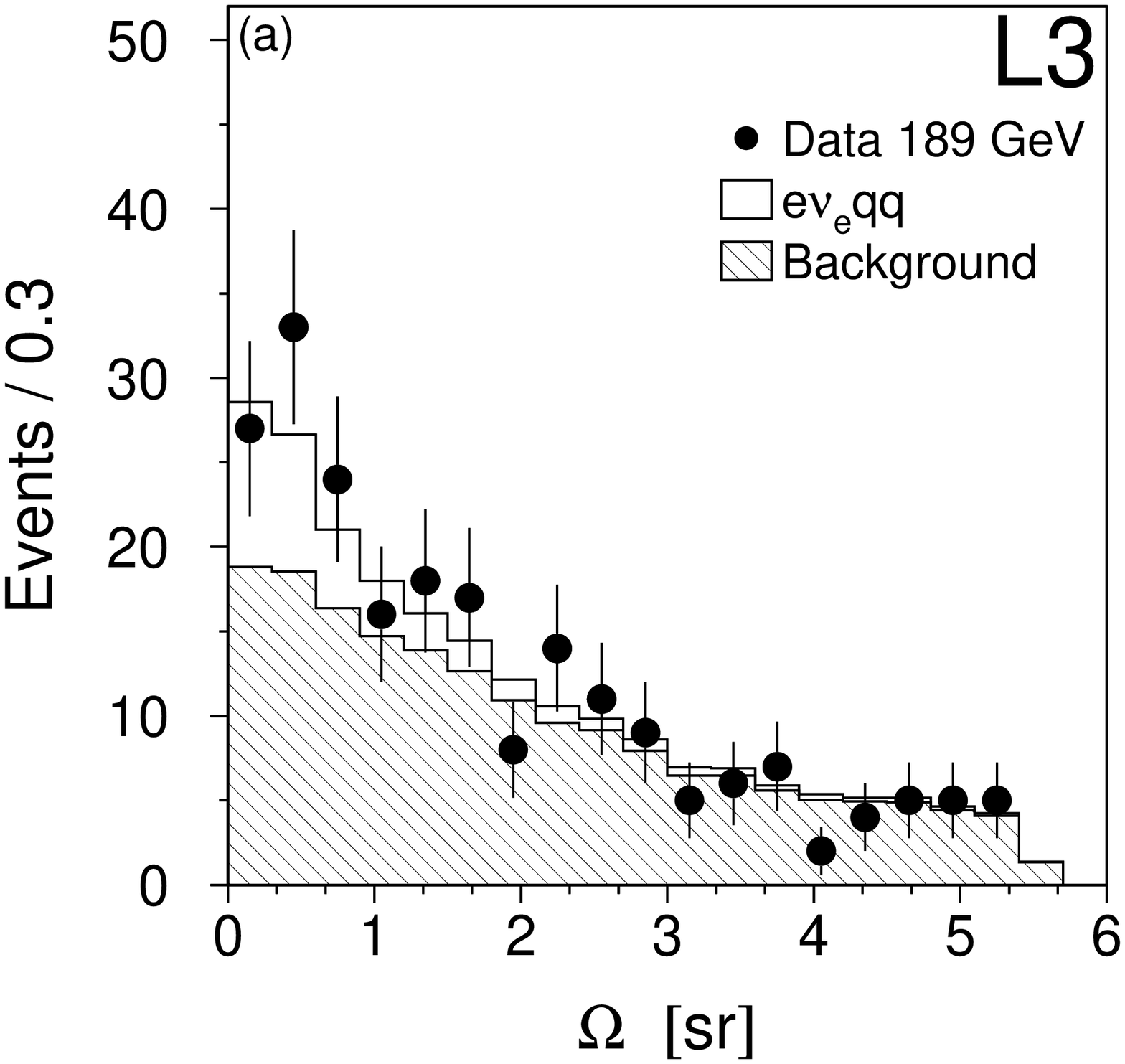,width=0.65\textwidth}}
    \mbox{\epsfig{file=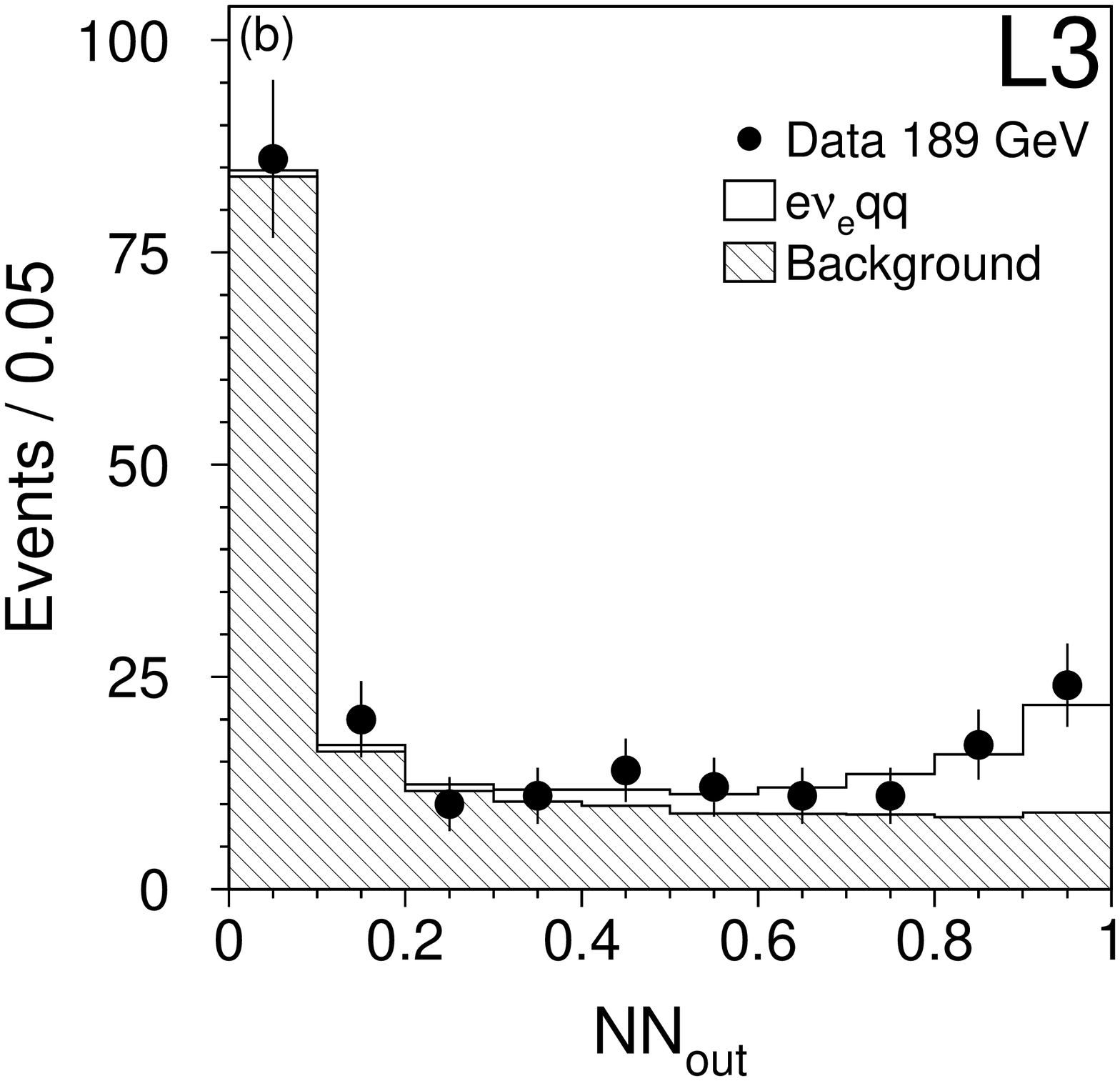,width=0.65\textwidth}}
  \end{center}
  \caption{\label{fig_om_nn}
           Distribution of (a) the 3-jet solid angle $\Omega$
           and (b) the neural network output for the selected
           hadronic events.
           The points are data, the hatched histograms represent the 
           background and the open histograms show the expected signal
           from $\mathrm{e} \nu_\mathrm{e} \q\q$ final states as 
           predicted by the EXCALIBUR Monte Carlo.}
\end{figure}

In order to differentiate further between the signal and the $\WW$
background, a discriminant variable $N\!N_{\rm out}$ is constructed
using a neural network. Nine variables are combined in
a feed-forward neural network~\cite{nnet}, with one hidden layer
and one output node. The input to the neural net includes the spherocity
of the event, the invariant mass of the two jets, the masses of the two
jets, the velocity $\beta $ of the hadronic system, the solid angle $\Omega$,
the resolution parameters $y_{23}$ and $y_{34}$ of the JADE jet finding
algorithm~\cite{jade} for which the number of jets in the event
changes from two to three and three to four, respectively, and the 
ratio of the mass to the energy of the least energetic jet
after forcing the event to three jets. As an example, the distribution of
the solid angle $\Omega$ for the selected events is shown in 
Figure~\ref{fig_om_nn}(a). The use of the neural network increases
the signal fraction in the selected sample to approximately 60\% for
large neural network output values, as shown in 
Figure~\ref{fig_om_nn}(b).

The cross-section of the process (1) for hadronic decays of the W
boson is determined by a binned log-likelihood fit to the neural
network output distribution of Figure~\ref{fig_om_nn}(b), assuming
Poisson statistics. The background contributions are fixed to the
corresponding Standard Model Monte Carlo predictions. The fitted
cross-section of the hadron channel is found to be:
\begin{center}
 $\sigma(\EE \to \mathrm{e} \nu_\mathrm{e} \q\q) = 
 0.41 \pm 0.11 \; \rm{pb}\,.$
\end{center}
A similar result is obtained if the GRC4F Monte Carlo is used for the
simulation of the signal. This result has to be compared with the expected
signal cross-section of $0.40\,\mbox{pb}$ predicted by EXCALIBUR and
$0.38\,\mbox{pb}$ predicted by GRC4F.

As a check of the analysis procedure, a fit of the total $\WW$
cross-section, keeping the single W contribution fixed to the EXCALIBUR
Monte Carlo prediction, gives the value $16.9 \pm 1.5 \; \rm{pb}$ in good
agreement with the Standard Model expectation of $16.7\,\mbox{pb}$ 
predicted by KORALW. The result of the fit with both $\WW$ and single W 
production cross-sections as free parameters is in good agreement with
the results above.


\subsection {Leptonic Final States}

The distinct feature of the process $\EE \to \Wen$, $\Wln$ is a high
energy lepton from W decay with no other significant activity in the
detector.

Events with one charged lepton (electron, muon or tau) with an energy
of at least $15\GeV{}$ are selected. The lepton identification is
based on the energy distribution in the electromagnetic and hadron
calorimeters associated with the trajectory of charged particles. The
total energy, $E_\mathrm{vis}$, is calculated as the sum of the lepton
energy, $E_\ell$, measured as discussed below, and the energies of all
neutral calorimetric clusters in the event. No other charged particle
activity is allowed.
The ratio $E_{\ell} / E_\mathrm{vis}$ is required to exceed
0.92 in order to suppress background from two-fermion production
$\EE \to \LL (\gamma)$. In addition, the energy in the
$\pm 0.22\,\mbox{rad}$ azimuthal angle sector along the missing
energy direction must be below $1\GeV{}$ ($0.2\GeV{}$ for muons).
The energy in the forward-backward luminosity calorimeters, $E_{\rm FB}$,
must be less than $70\GeV{}$. The accepted background is dominated by 
two-fermion production processes, especially radiative Bhabha scattering
in the case of the single electron final state, $\enu \ean$. Moreover,
significant contributions are due to 4-fermion final states that include
two neutrinos and fall outside the signal definition (\ref{eq:eqn02}).


\subsubsection {Single Electron Final States}

Electrons are identified as clusters in the electromagnetic calorimeter
consistent with an electromagnetic shower shape matched in azimuthal angle
with a track reconstructed in the central tracker.
For the single electron final states, the electron energy, $E_\mathrm{e}$,
measured in the electromagnetic calorimeter, must exceed
$20\GeV{}$ and the polar angle $\theta_\mathrm{e}$ must satisfy the condition
$| \cos\theta_\mathrm{e} | < 0.7$. This requirement reduces the contribution
from Bhabha and Compton scattering and from the process $\EE \to \EE
\nu \bar{\nu}$ where the $\EE$ pair originates from a low-mass virtual
photon. The $\EE \to \EE (\gamma)$ events constitute the dominant
contribution to the selected sample. 
The requirements $E_{\rm e} / E_\mathrm{vis} > 0.92$ and
$E_{\rm FB} < 70\GeV{}$ reduce significantly this contribution. The
acoplanarity angle between the direction of the electron and the most
energetic neutral cluster (if any) must be greater than 0.14~rad. 

A total of 8 events are observed with 11.9 expected from the Standard
Model including 7.9 events expected from the signal as predicted by the
EXCALIBUR Monte Carlo. The energy spectrum of the selected events is
presented in Figure~\ref{fig_lepton}(a).

The trigger efficiency is found to be 92\% from a control data sample.
The signal selection efficiency is estimated to be $(80 \pm 2)$\% using
the EXCALIBUR Monte Carlo program and $(78 \pm 3)$\% using GRC4F. The
major sources of the efficiency loss are due to the requirements  
$| \cos \theta_\mathrm{e} | < 0.7$ and $E_{\rm e} > 20\GeV$. A binned
log-likelihood fit to the electron energy spectrum results in:
\[
 \sigma(\EE \to \mathrm{e} \nu_\mathrm{e} \mathrm{e} \nu_\mathrm{e}) 
 = 0.021^{+0.026}_{-0.020} \; \rm{pb} \; ,
\]
to be compared with the signal cross-section of $0.056\,\mbox{pb}$
predicted by EXCALIBUR and $0.054\,\mbox{pb}$ predicted by GRC4F.


\subsubsection {Single Muon Final States}

Single muon final states are selected as events containing one
isolated muon, identified as a reconstructed track in the muon chambers 
and the central tracker. The muon energy, $E_\mu$, measured in the
muon chambers and in the central tracker, is required to exceed
$15~\GeV$. The fiducial volume for this analysis is defined to be
$|\cos\theta_\mu| < 0.85$. This latter requirement avoids significant
decrease of the trigger efficiency for single muons in the angular
region $|\cos\theta_\mu| > 0.85$. Within the acceptance, the trigger
efficiency exceeds 96\%. The total calorimetric energy not associated
to the muon must not exceed $3\GeV{}$. The rejection of background
from cosmic muons is based on the radial distance of closest approach
of the muon to the beam line and a match in azimuthal angle of the
muon chamber track with a track reconstructed in the central tracker.

A total of 10 events are observed with 9.8 expected from Standard
Model processes including 7.5 events from the signal. The energy spectrum
of the selected single muon candidates is presented in 
Figure~\ref{fig_lepton}(b). A signal efficiency of $(61 \pm 2)$\% is
estimated using the EXCALIBUR Monte Carlo program and $(56 \pm 2)$\% using
GRC4F. The main source of the efficiency loss is due to the geometrical
acceptance of the muon chambers. A binned log-likelihood fit to the muon
energy spectrum results in:
\[
 \sigma(\EE \to \mathrm{e} \nu_\mathrm{e} \mu \nu_\mu) 
 = 0.070^{+0.034}_{-0.027} \; \rm{pb}\; .
\]
The expected signal cross-section is $0.060\,\mbox{pb}$ according to
EXCALIBUR and $0.059\,\mbox{pb}$ according to GRC4F.


\subsubsection {Single Tau Final States}

Single tau final states are selected as events containing one low-multiplicity
hadronic jet. The calorimetric energy associated with the $\tau$ jet, $E_\tau$,
must exceed $15~\GeV$. The number of tracks reconstructed in the central
tracking system must be either 1 or 3.

A total of 4 events are observed with 3.6 expected from the Standard
Model processes including 2.1 events from the signal. The energy
spectrum of selected single tau candidates is presented in 
Figure~\ref{fig_lepton}(c). The signal efficiency is estimated to be
$(26 \pm 2)$\% using EXCALIBUR and $(29 \pm 2)$\% using GRC4F. The
trigger efficiency was studied and found to be in excess of 98\%.
A binned log-likelihood fit to the tau energy spectrum yields:
\[
 \sigma(\EE \to \mathrm{e} \nu_\mathrm{e} \tau \nu_\tau) 
 = 0.066^{+0.073}_{-0.050} \; \rm{pb} \; .
\]
The predictions for the signal cross-section are $0.060\,\mbox{pb}$
and $0.059\,\mbox{pb}$ according to EXCALIBUR and GRC4F, respectively.


\section {Systematic Uncertainties}

In case of the hadronic decay of the single W, the differences of the
EXCALIBUR and GRC4F signal modelling are taken into account in the
systematic uncertainty of the cross-section measurement. This
systematic uncertainty is found to be approximately 3\%.
In addition, the parameters describing the neural network structure
are varied and the analysis is repeated to allow an estimation of the
uncertainty due to the choice of the network, yielding a contribution
of 3\%. Compared to this, detector effects, studied by smearing and
shifting the kinematic variables that are fed into the network within
the experimental resolution, have a negligible impact on the result. 

In the lepton channel, the dominant systematic uncertainty of approximately
4\% arises from the signal modelling comparing the signal efficiencies
estimated using EXCALIBUR and GRC4F. The uncertainty due to the
identification of leptons is studied using control data samples of
two-fermion production and is found to be less than 1.5\%.

The uncertainty due to the Monte Carlo signal statistics ranges from 1\%
to 3\% on the cross-section depending on the decay channel. The systematic
uncertainty on the expected number of background events is essentially due
to the limited Monte Carlo statistics, with smaller contributions from the
uncertainties on the cross-sections and the selection efficiencies for the
background processes. The overall uncertainty on the total number of
background events ranges from 3\% to 4\% in the individual channels.
These uncertainties are
uncorrelated among individual channels and different centre-of-mass energies
and have negligible impact on the final results. Taking into account all the
contributions, the systematic uncertainty of the cross-section measurement
amounts to 5\% for the hadronic decay channel and 6\% overall.


\section {Results}

\subsection {Total Cross-Section}

\begin{figure} [ht]
  \begin{center}
    \mbox{\epsfig{file=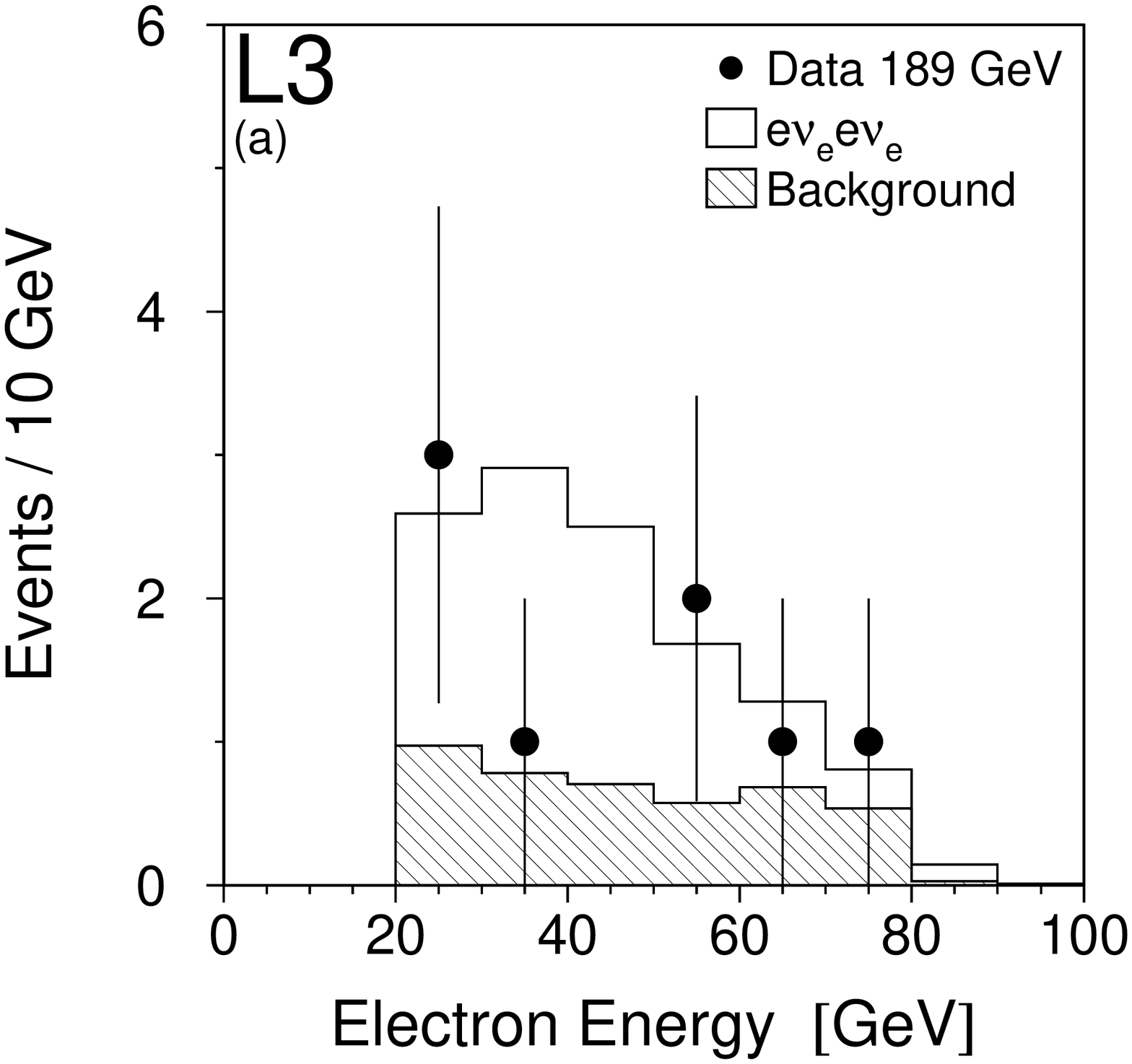,width=0.49\textwidth}}
    \mbox{\epsfig{file=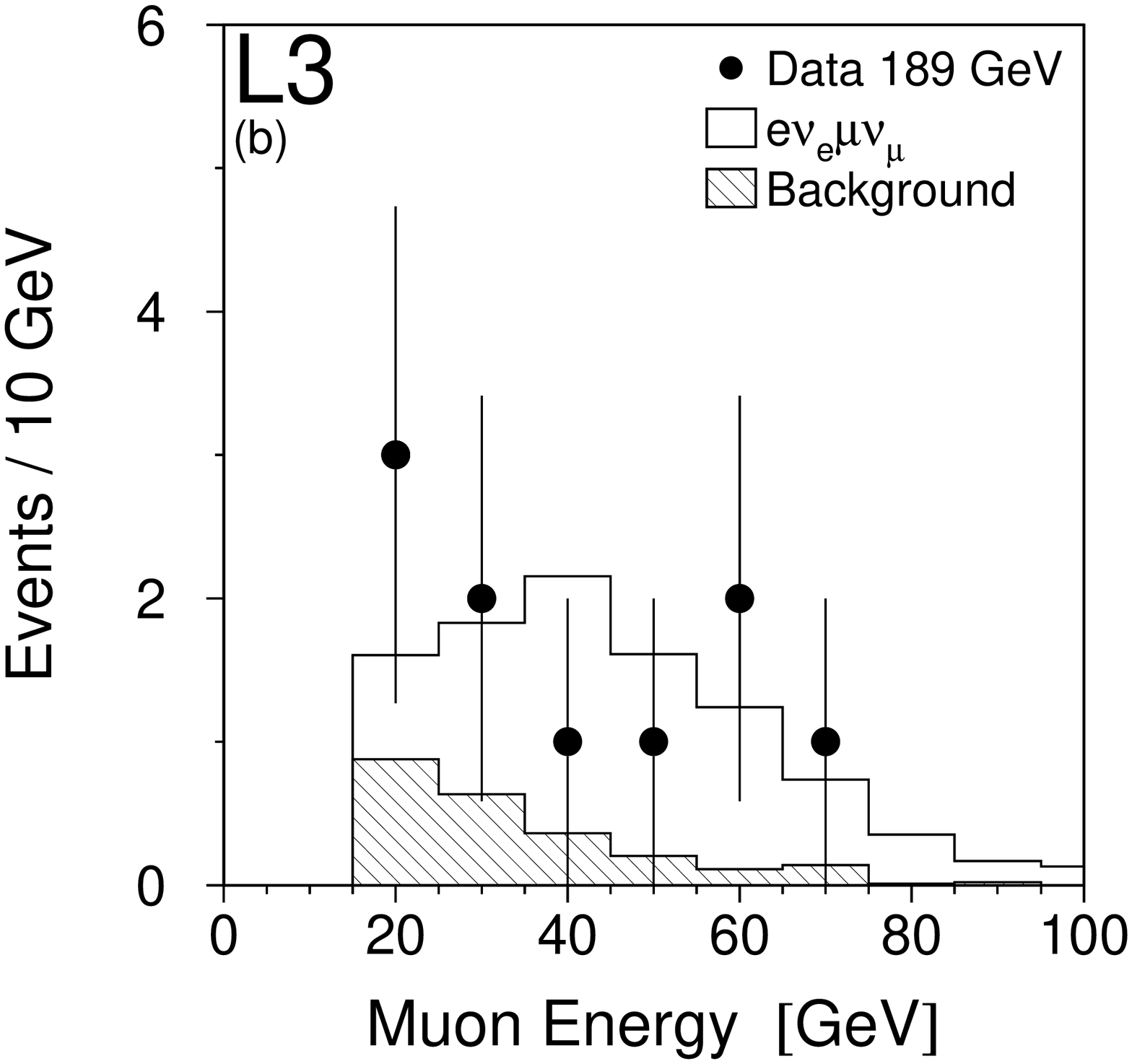,width=0.49\textwidth}}
    \mbox{\epsfig{file=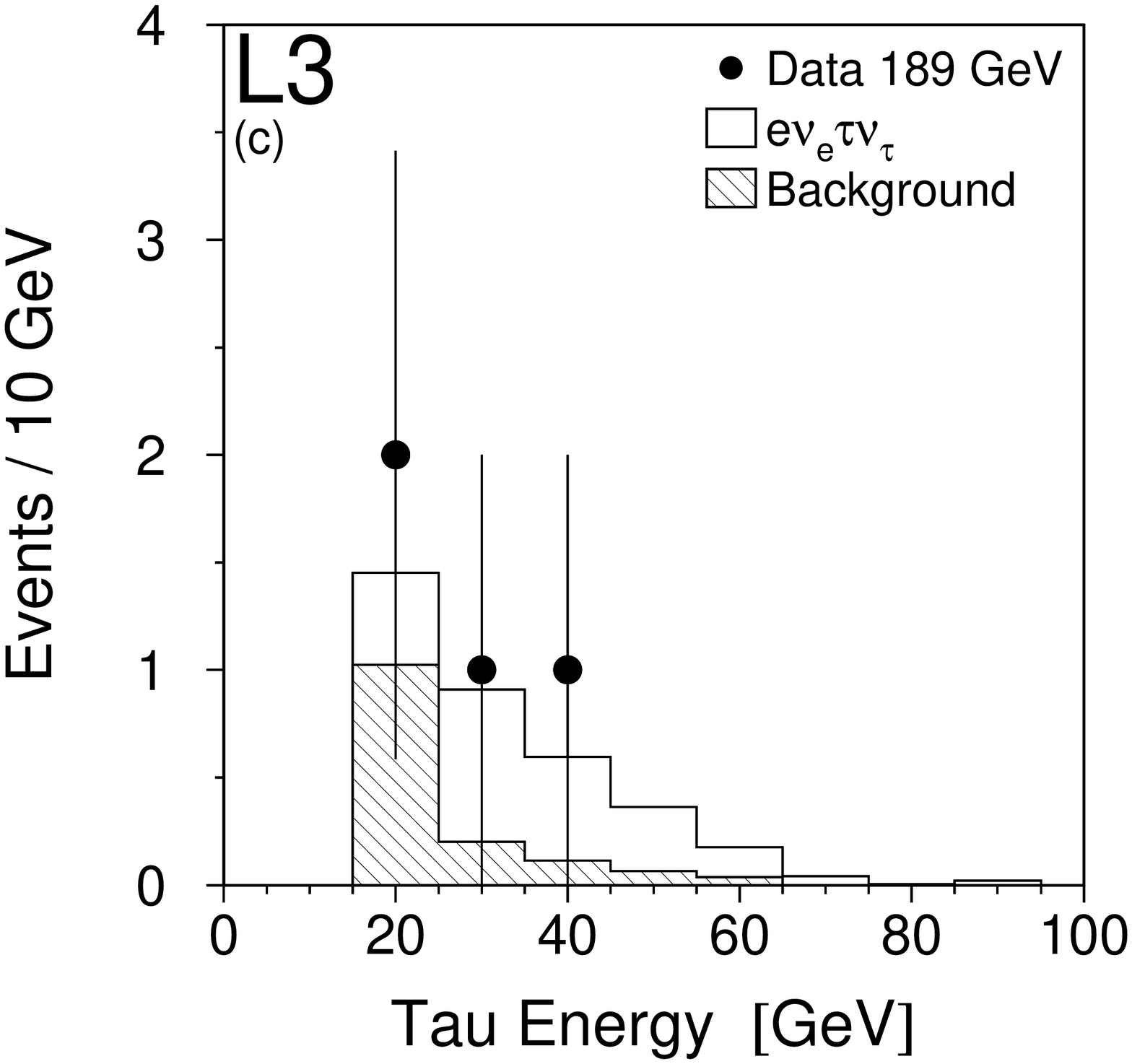,width=0.49\textwidth}}
    \mbox{\epsfig{file=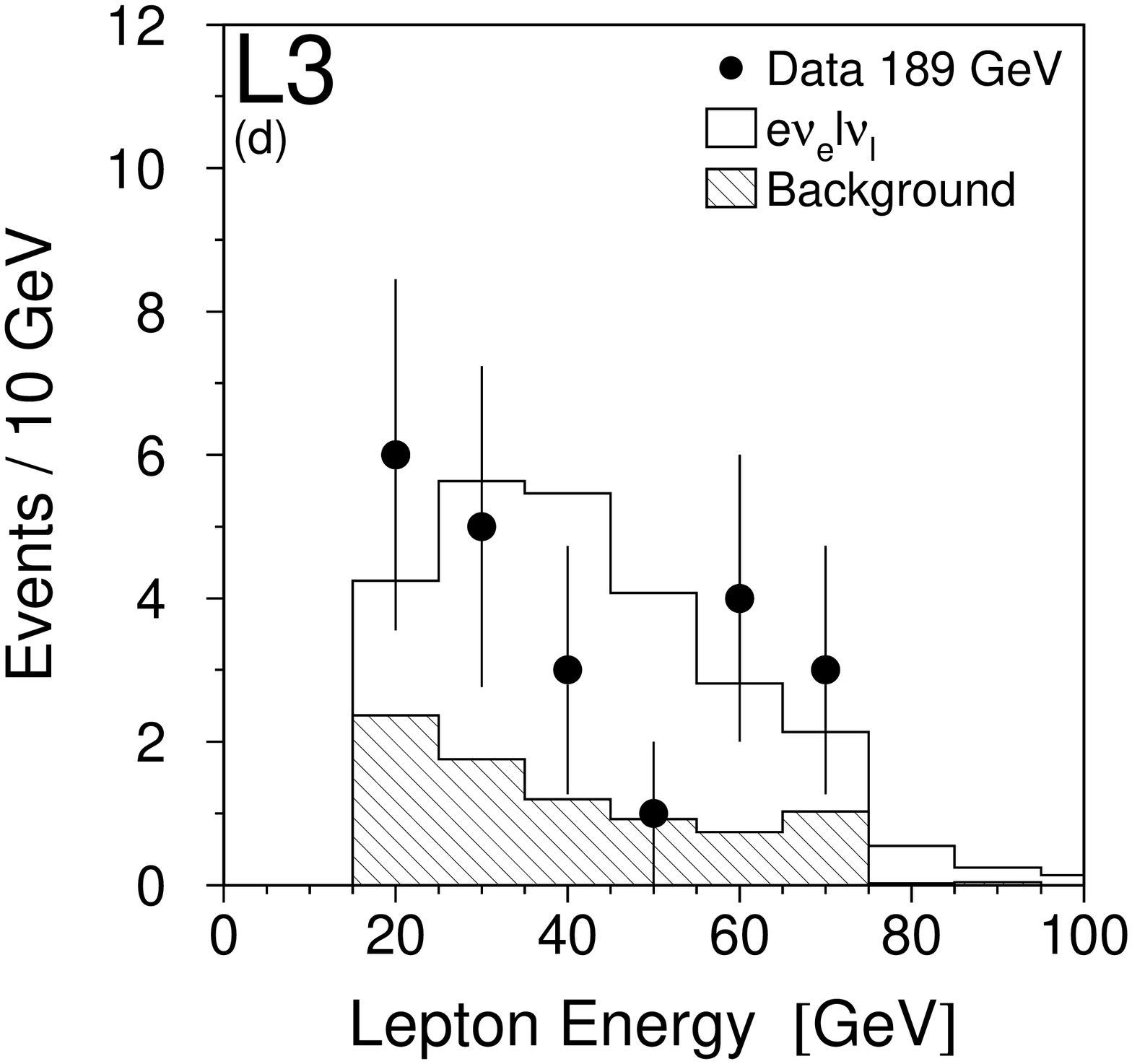,width=0.49\textwidth}}
  \end{center}
  \caption{\label{fig_lepton}
           The energy spectra of the selected (a) single electron,
           (b) single muon, and (c) single tau candidates. The 
           sum of the different lepton energy spectra is shown 
           in (d). The points are data, the hatched histograms
           correspond to the background contribution. The open
           histograms show the expected signal from
           $\mathrm{e} \nu_\mathrm{e} \ell \nu_\ell$ final states
           as predicted by the EXCALIBUR Monte Carlo.}
\end{figure}

\begin{figure} [ht]
  \begin{center}
    \mbox{\epsfig{file=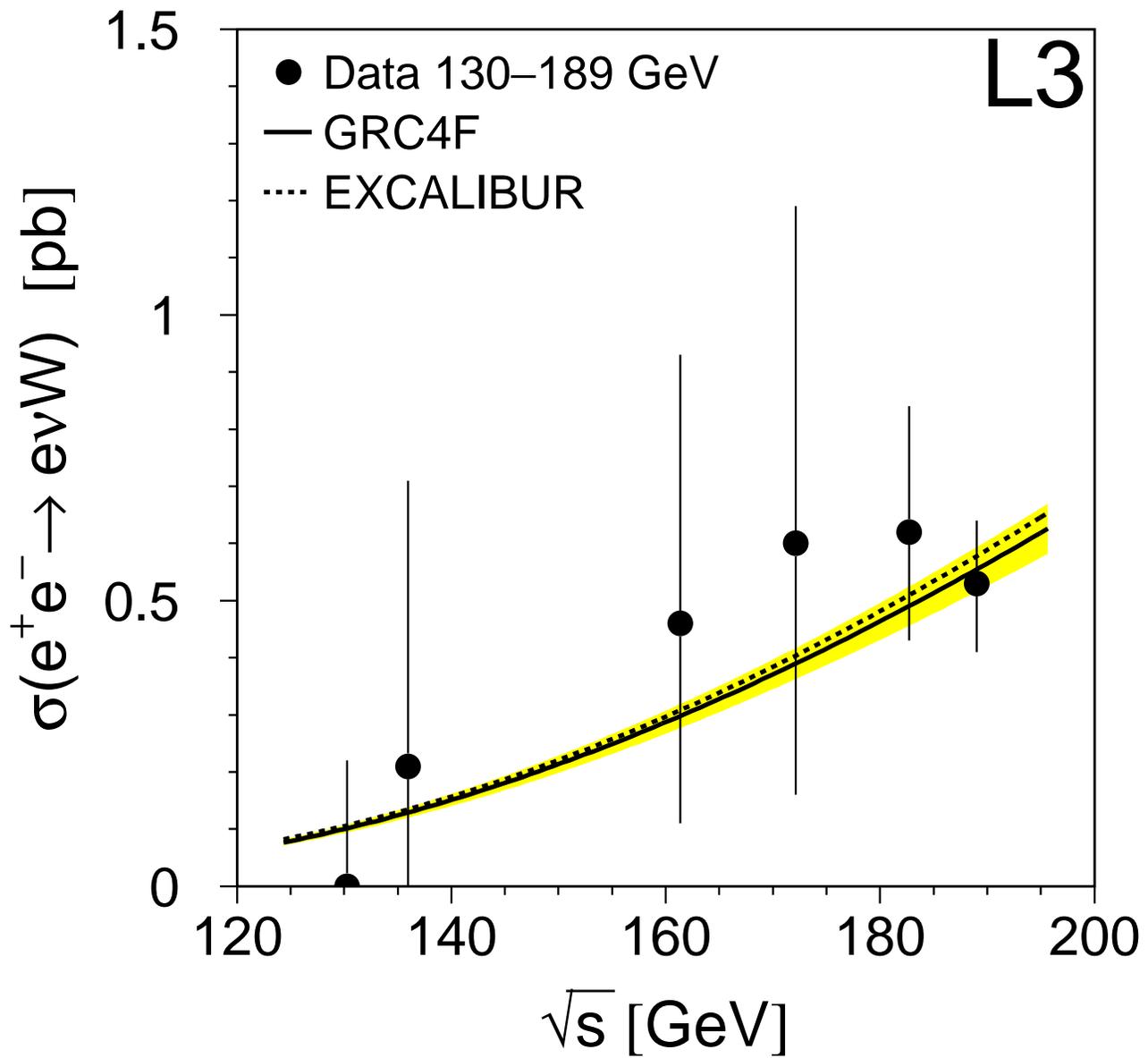,width=1.0\textwidth}}
  \end{center}
  \caption{\label{fig_xs}
           The measured cross-section of single W production within our
           phase-space cuts as a function of the centre-of-mass energy.
           The solid and dashed lines show predictions of the GRC4F
           and EXCALIBUR Monte Carlo programs, respectively. The estimated
           theoretical uncertainty of $\pm 7\%$ is indicated by the band.}
\end{figure}

The total cross-section of single W production is determined from a
binned likelihood fit to the distributions of the neural network
output presented in Figure~\ref{fig_om_nn}(b) and the lepton energy
spectra shown in Figures~\ref{fig_lepton}(a) -- (c). The sum of the
different lepton energy distributions is presented in 
Figure~\ref{fig_lepton}(d).
The background shapes and normalisations are fixed to the Monte Carlo
prediction. The fitted signal cross-section, $\sigma(\EE \to \enw)$,
corresponds to that of the process $\rm \EE \to e \nu_e {\rm f\/f}$, where
f\/f denotes a sum of all $\rm \ell \nu_\ell$ and $\q\q$ final
states satisfying the phase-space conditions (2). The total single W
boson cross-section at $\sqrt{s}=189\GeV$ is then determined to be:
\[
 \sigma(\EE \to \mathrm{e} \nu_\mathrm{e} {\rm W}) 
 = 0.53 \pm 0.12 \pm 0.03 \; \rm{pb} \,,
\]
where the first error is statistical and the second systematic. The
measured cross-section value is consistent with the Standard Model
prediction of $0.57\,\mbox{pb}$ calculated with EXCALIBUR and
$0.56\,\mbox{pb}$ calculated with GRC4F. The dependence 
of the cross-section on the centre-of-mass energy agrees well
with the Monte Carlo predictions as shown in Figure~\ref{fig_xs},
including our previous measurements at centre-of-mass energies
between $130\GeV$ and $183\GeV$~\cite{l3wen2}.


\subsection {\boldmath$\rm W W \gamma$ Gauge Couplings}

The electromagnetic gauge couplings $\rm \kappa_\gamma$ and $\rm
\lambda_\gamma$ describing the WW$\gamma$ vertex are determined from a
binned maximum-likelihood fit similar to the one used for the cross
section determination. In the fit each Monte Carlo event is assigned a
weight that depends on the generated 4-fermion event kinematics and
the values of the gauge couplings $\rm \kappa_\gamma$ and
$\rm \lambda_\gamma$.
The dependence of the background from $\WW$ production on
the gauge couplings is
also taken into account. The weight is calculated using the matrix
element as implemented in EXCALIBUR, imposing constraints on the
triple gauge boson couplings $\kappa_\mathrm{Z}$ and
$\lambda_\mathrm{Z}$ arising from the $SU(2) \times U(1)$ gauge
invariance: $\kappa_\mathrm{Z} = g_{1}^{\mathrm{Z}} - \tan^2
\theta_{\rm w} (\kappa_\gamma - 1)$ and $\lambda_\mathrm{Z} =
\lambda_\gamma$. These constraints affect only the background
contributions, as the signal process depends on $\rm \lambda_\gamma$
and $\rm \kappa_\gamma$ only. The general analysis of
the remaining C- and P-conserving couplings,
$g_{1}^{\mathrm{Z}}$, $\lambda_\gamma$ and $\kappa_\gamma$,
can only be done with full consideration of the $\WW$ production. In
the present analysis we fix the weak charge of the W bosons to its 
Standard Model value, $g_{1}^{\mathrm{Z}} = 1$, and focus on the
electromagnetic properties of W bosons.

The dependence of the coupling determination on the total cross-section
for single W boson production is tested repeating the likelihood fit with
a $\pm 7\%$ variation of the signal cross-section.
The corresponding systematic uncertainty is $\pm 0.04$ for $\kappa_\gamma$
and $\pm 0.02$ for $\lambda_\gamma$. 
Comparing the signal description of the two Monte Carlo Generators, EXCALIBUR
and GRC4F, shows an additional systematic uncertainty of $\pm 0.05$ and
$\pm 0.06$
on $\kappa_\gamma$ and $\lambda_\gamma$, respectively.
The agreement between the two generators for various anomalous couplings
is checked. No coupling-dependence of the ratio of the
two cross-section predictions was found.

In addition, the cross-section of the background contributions coming
from $\WW$ and ZZ production are varied to allow an estimation of the
systematic uncertainty. The influence of these uncertainties on the
coupling determination is found to be less than $\pm 0.02$. 

The estimated systematic uncertainty is assumed to be Gaussian and fully
correlated between individual channels. The systematic uncertainty
in the cross-section determination is taken into account by convolution of
the likelihood function with a Gaussian in the fit.

\begin{figure} [ht]
  \begin{center}
    \mbox{\epsfig{file=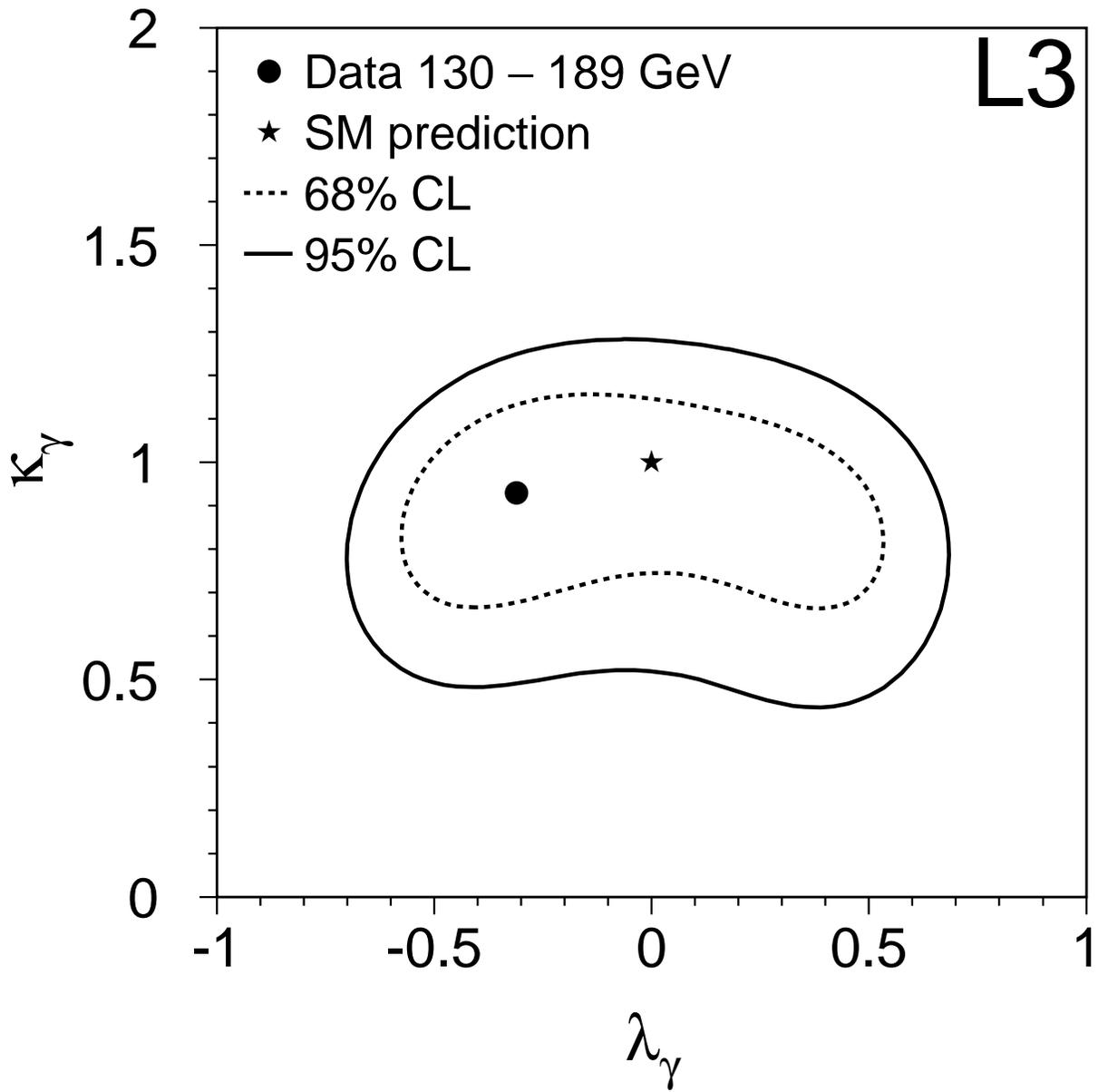,width=1.0\textwidth}}
  \end{center}
  \caption{\label{fig_cn}
           The contours corresponding to 68\% and 95\% confidence level
           in the $\rm \kappa_\gamma - \rm \lambda_\gamma$ plane. The
           point indicates the global minimum from the 2-parameter fit,
           to be compared with the Standard Model prediction indicated
           by the star.}
\end{figure}

For the fit to the couplings, we combine the single W data at $\sqrt{s} =
189\GeV$ presented here with our single W data already 
published~\cite{l3wen2}. Single W boson production is particularly
sensitive to the gauge coupling $\kappa_\gamma$. Thus, this coupling
is determined in a fit fixing $\lambda_\gamma = 0$ to be:
\begin{eqnarray*}
  \kappa_\gamma  & = & +0.96^{+0.15}_{-0.17} \pm 0.09 \, ,
\end{eqnarray*}
where the first error is statistical and the second systematic. Fixing
$\kappa_\gamma = 1$ and performing a fit for $\lambda_\gamma$ yields:
\begin{eqnarray*}
  \lambda_\gamma & = & -0.26^{+0.53}_{-0.19} \pm 0.13 \, .
\end{eqnarray*}
Varying both couplings $\lambda_\gamma$ and $\kappa_\gamma$ freely in the
fit yields:
\begin{center}
\begin{tabular}{rcr@{}l@{}l}
  $\kappa_\gamma$  & = & $+0.93$ & $\pm 0.16$            & $\pm 0.09$ \\[0.7ex]
  $\lambda_\gamma$ & = & $-0.31$ & ${}^{+0.68}_{-0.19}$ & $\pm 0.13$ \,,
\end{tabular}
\end{center}
with a correlation coefficient of $+37\%$. The 68\% and 95\% confidence level
contours on $\kappa_\gamma$ and $\lambda_\gamma$ are shown in 
Figure~\ref{fig_cn}. The results are consistent
with the absence of anomalous contributions to WW$\gamma$ couplings. 
The limits on $\kappa_\gamma$ and $\lambda_\gamma$ at 95\% confidence 
level are:
\begin{eqnarray*}
  0.56 \; < &  \kappa_\gamma & < \; 1.29 
  \quad \mbox{for} \quad \lambda_\gamma = 0 \\
 -0.67 \; < & \lambda_\gamma & < \; 0.59 
  \quad \mbox{for} \quad \kappa_\gamma = 1  \,. 
\end{eqnarray*}
These results represent a major improvement in the accuracy on
the triple gauge boson couplings $\kappa_\gamma$ and $\lambda_\gamma$
compared to our previous publications on single W boson 
production~\cite{l3wen1,l3wen2}. They are complementary to
measurements based on $\WW$ production at LEP~\cite{delphiwen,lepww}
or determined at the Tevatron $\mathrm{p}\overline{\mathrm{p}}$ 
collider~\cite{tevatron}.


\section*{Acknowledgements}
 
We wish to express our gratitude to the CERN accelerator divisions for
the excellent performance of the LEP machine. We acknowledge the
efforts of the engineers, technicians and support staff who have
participated in the construction and maintenance of this experiment.


\newpage
\section*{Author List}
\typeout{   }     
\typeout{Using author list for paper 212 only}
\typeout{$Modified: Tue May  17 13:45:26 2000 by bobbink $}
\typeout{!!!!  This should only be used with document option a4p!!!!}
\typeout{   }
%
%
%
%
%
%

\newcount\tutecount  \tutecount=0
\def\tutenum#1{\global\advance\tutecount by 1 \xdef#1{\the\tutecount}}
\def\tute#1{$^{#1}$}
\tutenum\aachen            
\tutenum\nikhef            
\tutenum\mich              
\tutenum\lapp              
\tutenum\basel             
\tutenum\lsu               
\tutenum\beijing           
\tutenum\berlin            
\tutenum\bologna           
\tutenum\tata              
\tutenum\ne                
\tutenum\bucharest         
\tutenum\budapest          
\tutenum\mit               
\tutenum\debrecen          
\tutenum\florence          
\tutenum\cern              
\tutenum\wl                
\tutenum\geneva            
\tutenum\hefei             
\tutenum\seft              
\tutenum\lausanne          
\tutenum\lecce             
\tutenum\lyon              
\tutenum\madrid            
\tutenum\milan             
\tutenum\moscow            
\tutenum\naples            
\tutenum\cyprus            
\tutenum\nymegen           
\tutenum\caltech           
\tutenum\perugia           
\tutenum\cmu               
\tutenum\prince            
\tutenum\rome              
\tutenum\peters            
\tutenum\potenza           
\tutenum\salerno           
\tutenum\ucsd              
\tutenum\santiago          
\tutenum\sofia             
\tutenum\korea             
\tutenum\alabama           
\tutenum\utrecht           
\tutenum\purdue            
\tutenum\psinst            
\tutenum\zeuthen           
\tutenum\eth               
\tutenum\hamburg           
\tutenum\taiwan            
\tutenum\tsinghua          

{
\parskip=0pt
\noindent
{\bf The L3 Collaboration:}
\ifx\selectfont\undefined
 \baselineskip=10.8pt
 \baselineskip\baselinestretch\baselineskip
 \normalbaselineskip\baselineskip
 \ixpt
\else
 \fontsize{9}{10.8pt}\selectfont
\fi
\medskip
\tolerance=10000
\hbadness=5000
\raggedright
\hsize=162truemm\hoffset=0mm
\def\r{\rlap,}
\noindent

M.Acciarri\r\tute\milan\
P.Achard\r\tute\geneva\ 
O.Adriani\r\tute{\florence}\ 
M.Aguilar-Benitez\r\tute\madrid\ 
J.Alcaraz\r\tute\madrid\ 
G.Alemanni\r\tute\lausanne\
J.Allaby\r\tute\cern\
A.Aloisio\r\tute\naples\ 
M.G.Alviggi\r\tute\naples\
G.Ambrosi\r\tute\geneva\
% ...
\newpage
\begin{list}{A}{\itemsep=0pt plus 0pt minus 0pt\parsep=0pt plus 0pt minus 0pt
                \topsep=0pt plus 0pt minus 0pt}
\item[\aachen]
 I. Physikalisches Institut, RWTH, D-52056 Aachen, FRG$^{\S}$\\
 III. Physikalisches Institut, RWTH, D-52056 Aachen, FRG$^{\S}$
\item[\nikhef] National Institute for High Energy Physics, NIKHEF, 
     and University of Amsterdam, NL-1009 DB Amsterdam, The Netherlands
\item[\mich] University of Michigan, Ann Arbor, MI 48109, USA
\item[\lapp] Laboratoire d'Annecy-le-Vieux de Physique des Particules, 
     LAPP,IN2P3-CNRS, BP 110, F-74941 Annecy-le-Vieux CEDEX, France
\item[\basel] Institute of Physics, University of Basel, CH-4056 Basel,
     Switzerland
\item[\lsu] Louisiana State University, Baton Rouge, LA 70803, USA
\item[\beijing] Institute of High Energy Physics, IHEP, 
  100039 Beijing, China$^{\triangle}$ 
\item[\berlin] Humboldt University, D-10099 Berlin, FRG$^{\S}$
\item[\bologna] University of Bologna and INFN-Sezione di Bologna, 
     I-40126 Bologna, Italy
\item[\tata] Tata Institute of Fundamental Research, Bombay 400 005, India
\item[\ne] Northeastern University, Boston, MA 02115, USA
\item[\bucharest] Institute of Atomic Physics and University of Bucharest,
     R-76900 Bucharest, Romania
\item[\budapest] Central Research Institute for Physics of the 
     Hungarian Academy of Sciences, H-1525 Budapest 114, Hungary$^{\ddag}$
\item[\mit] Massachusetts Institute of Technology, Cambridge, MA 02139, USA
\item[\debrecen] KLTE-ATOMKI, H-4010 Debrecen, Hungary$^\P$
\item[\florence] INFN Sezione di Firenze and University of Florence, 
     I-50125 Florence, Italy
\item[\cern] European Laboratory for Particle Physics, CERN, 
     CH-1211 Geneva 23, Switzerland
\item[\wl] World Laboratory, FBLJA  Project, CH-1211 Geneva 23, Switzerland
\item[\geneva] University of Geneva, CH-1211 Geneva 4, Switzerland
\item[\hefei] Chinese University of Science and Technology, USTC,
      Hefei, Anhui 230 029, China$^{\triangle}$
\item[\seft] SEFT, Research Institute for High Energy Physics, P.O. Box 9,
      SF-00014 Helsinki, Finland
\item[\lausanne] University of Lausanne, CH-1015 Lausanne, Switzerland
\item[\lecce] INFN-Sezione di Lecce and Universit\'a Degli Studi di Lecce,
     I-73100 Lecce, Italy
\item[\lyon] Institut de Physique Nucl\'eaire de Lyon, 
     IN2P3-CNRS,Universit\'e Claude Bernard, 
     F-69622 Villeurbanne, France
\item[\madrid] Centro de Investigaciones Energ{\'e}ticas, 
     Medioambientales y Tecnolog{\'\i}cas, CIEMAT, E-28040 Madrid,
     Spain${\flat}$ 
\item[\milan] INFN-Sezione di Milano, I-20133 Milan, Italy
\item[\moscow] Institute of Theoretical and Experimental Physics, ITEP, 
     Moscow, Russia
\item[\naples] INFN-Sezione di Napoli and University of Naples, 
     I-80125 Naples, Italy
\item[\cyprus] Department of Natural Sciences, University of Cyprus,
     Nicosia, Cyprus
\item[\nymegen] University of Nijmegen and NIKHEF, 
     NL-6525 ED Nijmegen, The Netherlands
\item[\caltech] California Institute of Technology, Pasadena, CA 91125, USA
\item[\perugia] INFN-Sezione di Perugia and Universit\'a Degli 
     Studi di Perugia, I-06100 Perugia, Italy   
\item[\cmu] Carnegie Mellon University, Pittsburgh, PA 15213, USA
\item[\prince] Princeton University, Princeton, NJ 08544, USA
\item[\rome] INFN-Sezione di Roma and University of Rome, ``La Sapienza",
     I-00185 Rome, Italy
\item[\peters] Nuclear Physics Institute, St. Petersburg, Russia
\item[\potenza] INFN-Sezione di Napoli and University of Potenza, 
     I-85100 Potenza, Italy
\item[\salerno] University and INFN, Salerno, I-84100 Salerno, Italy
\item[\ucsd] University of California, San Diego, CA 92093, USA
\item[\santiago] Dept. de Fisica de Particulas Elementales, Univ. de Santiago,
     E-15706 Santiago de Compostela, Spain
\item[\sofia] Bulgarian Academy of Sciences, Central Lab.~of 
     Mechatronics and Instrumentation, BU-1113 Sofia, Bulgaria
\item[\korea]  Laboratory of High Energy Physics, 
     Kyungpook National University, 702-701 Taegu, Republic of Korea
\item[\alabama] University of Alabama, Tuscaloosa, AL 35486, USA
\item[\utrecht] Utrecht University and NIKHEF, NL-3584 CB Utrecht, 
     The Netherlands
\item[\purdue] Purdue University, West Lafayette, IN 47907, USA
\item[\psinst] Paul Scherrer Institut, PSI, CH-5232 Villigen, Switzerland
\item[\zeuthen] DESY, D-15738 Zeuthen, 
     FRG
\item[\eth] Eidgen\"ossische Technische Hochschule, ETH Z\"urich,
     CH-8093 Z\"urich, Switzerland
\item[\hamburg] University of Hamburg, D-22761 Hamburg, FRG
\item[\taiwan] National Central University, Chung-Li, Taiwan, China
\item[\tsinghua] Department of Physics, National Tsing Hua University,
      Taiwan, China
\item[\S]  Supported by the German Bundesministerium 
        f\"ur Bildung, Wissenschaft, Forschung und Technologie
\item[\ddag] Supported by the Hungarian OTKA fund under contract
numbers T019181, F023259 and T024011.
\item[\P] Also supported by the Hungarian OTKA fund under contract
  numbers T22238 and T026178.
\item[$\flat$] Supported also by the Comisi\'on Interministerial de Ciencia y 
        Tecnolog{\'\i}a.
\item[$\sharp$] Also supported by CONICET and Universidad Nacional de La Plata,
        CC 67, 1900 La Plata, Argentina.
\item[$\diamondsuit$] Also supported by Panjab University, Chandigarh-160014, 
        India.
\item[$\triangle$] Supported by the National Natural Science
  Foundation of China.
\end{list}
}
\vfill




\clearpage
\vfill

\clearpage
\vfill

\clearpage
\vfill

\clearpage
\vfill

\end{document}